\documentclass[aps,prresearch,reprint,superscriptaddress,amsmath,amssymb]{revtex4-2}
\usepackage[utf8]{inputenc}
\usepackage[dvipsnames]{xcolor}
\usepackage{amsmath}
\usepackage{graphicx}
\usepackage[dvipsnames]{xcolor}
\definecolor{citeblue}{RGB}{25,60,120}
\definecolor{urlteal}{RGB}{0,100,105}
\usepackage{hyperref}
\hypersetup{
  bookmarks=false,
  breaklinks=false,
  backref=false,
  colorlinks=true,
  linkcolor=BrickRed,
  citecolor=citeblue,
  urlcolor=urlteal
}

\makeatletter
\usepackage{bm}
\usepackage[normalem]{ulem}

\usepackage{xcolor}

\makeatother

\begin{document}
\global\long\def\vect#1{\overrightarrow{\mathbf{#1}}}%

\global\long\def\abs#1{\left|#1\right|}%

\global\long\def\av#1{\left\langle #1\right\rangle }%

\global\long\def\imp#1{\left\langle #1\right\rangle _{\text{imp}}}%

\global\long\def\ket#1{\left|#1\right\rangle }%

\global\long\def\bra#1{\left\langle #1\right|}%

\global\long\def\tensorproduct{\otimes}%

\global\long\def\braket#1#2{\left\langle #1\mid#2\right\rangle }%

\global\long\def\bb#1{\boldsymbol{#1}}%

\global\long\def\b#1{\mathbf{#1}}%

\global\long\def\cal#1{\mathcal{#1}}%

\global\long\def\scr#1{\mathscr{#1}}%

\global\long\def\norm#1{\mathit{#1}}%

\global\long\def\braket#1#2{\left\langle #1|#2\right\rangle }%

\global\long\def\p#1{\left(#1\right)}%

\global\long\def\t#1{\text{#1}}%

\global\long\def\s#1{{\displaystyle {\displaystyle #1}}}%

\global\long\def\com#1#2{\left[#1,#2\right]}%

\global\long\def\an#1#2{\left\{  #1,#2\right\}  }%

\global\long\def\omv{\overrightarrow{\Omega}}%

\global\long\def\inf{\infty}%

\title{Local orbital magnetization and Chern markers from equilibrium local currents}
\author{Henrique P. Veiga}
\email{henriqueveigacj@gmail.com}
\affiliation{Centro de Física das Universidades do Minho e do Porto, LaPMET, Departamento
de Física e Astronomia, Faculdade de Ciências, Universidade do Porto,
4169-007 Porto, Portugal}
\affiliation{School of Physics, Engineering and Technology and York Centre for Quantum Technologies, University of York, YO10 5DD, York, United Kingdom}

\author{Aires Ferreira}
\affiliation{School of Physics, Engineering and Technology and York Centre for Quantum Technologies, University of York, YO10 5DD, York, United Kingdom}

\author{Caio H. Lewenkopf}
\affiliation{Instituto de F\'isica,
Universidade Federal do Rio de Janeiro,
21941-972 Rio de Janeiro, RJ, Brazil}
\affiliation{Institut für Theoretische Physik,
Universität Regensburg,
D-93040 Regensburg, Germany}

\author{João M. Viana Parente Lopes}
\affiliation{Centro de Física das Universidades do Minho e do Porto, LaPMET, Departamento
de Física e Astronomia, Faculdade de Ciências, Universidade do Porto,
4169-007 Porto, Portugal}

\begin{abstract}
Local currents carry the orbital magnetization of a macroscopic sample and are key for modeling inhomogeneous magnetic materials and for understanding magnetoelectric and topological responses. However,
producing realistic simulated images of local currents in inhomogeneous
systems remains challenging. Building upon recent work on stochastic
estimation methods for tight-binding models, we show that all matrix elements of positive semidefinite
operators defined on a lattice can be obtained at once from a single set of filtered random
vectors.
This eliminates the need for independent simulations previously required for off-diagonal elements, which significantly speeds up the calculation of spatial maps of local currents. Further, we express  bond currents in terms of loop currents that obey a discrete Poisson equation, with the current vorticity acting as the source term.
Its solution yields a local orbital magnetization.
In analogy with the global Středa relation, the derivative of this magnetization with respect to the chemical potential defines a new local Chern marker. Built entirely from bond currents, it requires no position operator and is not bound by the zero-sum rule of projector markers in open samples. Hence, it is free of boundary-compensation artifacts, and its sample average converges to the bulk Chern number in a controlled fashion.
\end{abstract}
\maketitle

\section{Introduction}
\label{sec:intro}

Spatially resolved local observables reveal physical information that global  measurements inevitably average away.
In 2D crystals, chiral edge channels~\cite{uri2020}, current vortices around defects~\cite{aharonsteinberg2022,palm2024} and localization transitions~\cite{zhang2009, veiga2026markov} are invisible to net current
measurements.
Scanning magnetometry has recently proven its suitability to directly probe these currents, and the magnetization they generate, with fine spatial resolution~\cite{nowack2013,uri2020,tschirhart2021}, down to the moiré scale, where orbital ferromagnets break up into domains of opposite Chern number~\cite{grover2022}. Scanning tunneling spectroscopy has also been proposed as a probe of loop-current order, which produces weak local magnetism in cuprates, kagome metals, and at the surface of Sr$_{2}$RuO$_{4}$~\cite{morisseau2026}.

Orbital magnetization links equilibrium local currents to topological phases~\cite{macdonald1984,thonhauser2005,ceresoli2006, Lima2022}. These observables are divergence-free and carry no net transport, so by the Helmholtz decomposition they manifest as the curl of a magnetization density.
The Středa relation$\,$\cite{streda1982} links the chemical-potential derivative of the global magnetization to the Hall conductivity, which in insulators is quantized by the Chern number.
This invariant is defined as a Brillouin-zone integral of the Berry
curvature. Without translation symmetry, crystal momentum is no longer a good quantum
number,  requiring the Chern number to
be defined in real space.
Local Chern markers$\,$\cite{kitaev2006,bianco2011,varjas2020} have become the standard tool for probing topology in real space, enabling the resolution of Chern mosaics in quasi-periodic systems$\,$\cite{grover2022,antao2026} and  the investigation of
disorder-driven topological transitions$\,$\cite{favata2025, Assuncao2025, kovalenka2026,ding2026}.
However, they rely on the position operator, and in a finite sample with open boundaries the marker, being the trace of a commutator of finite matrices, sums identically to zero.
This behavior is guaranteed by
a boundary contribution that grows with the system size and exactly compensates the bulk value. Markers and other topological indicators \cite{dornellas2022,markov2024,Setescac2025,repse2026,cioni2026} that avoid this compensation have recently been proposed. In parallel, the local orbital magnetization has been formulated as a nearsighted property of the ground-state density matrix~\cite{bianco2013}, as the field whose curl generates the
equilibrium current~\cite{daido2026}, and through the field response of the local density of
states~\cite{saati2025}. On a lattice, the magnetization density of a plaquette is the response
to a flux threaded through it, and differences of these densities give the bond
currents~\cite{nathan2017quantized}.

Comparing local markers with scanning-probe images demands samples large enough to approach the thermodynamic limit while capturing the disorder  present in experiments \cite{lachman2015,marguerite2019,uri2020twist,grover2022,ferguson2023}.
Since both local currents and topological markers are matrix elements of functions of the Hamiltonian, Chebyshev expansions \cite{weisse2006,Ferreira2015,garcia2015,fan2021linear,joao2020kite,joao2020nonlinear,santospires2020landauer,veiga2024unambiguous,Pires2022,Castro_2024} serve as the natural tool, and have recently been applied to the global orbital magnetization of large
disordered samples \cite{vidarte2026}.
In Ref.~\cite{veiga2026markov}, all diagonal elements of a spectral operator were sampled at once with a positive semidefinite estimator whose fluctuations are bounded by the Markov inequality.
Off-diagonal elements, however, necessitate rotating disjoint pairs of sites, at the cost of two simulations (real and imaginary parts) per tiling of the lattice into pairs.

The purpose of this paper is twofold. First, we introduce an algorithmic contribution to spectral methods:
We show that a single set of filtered random vectors gives every off-diagonal element of a spectral operator.
This removes the multiplicative cost of the tilings without computational time or memory overhead.
Consequently, a complete map of local currents requires a single Chebyshev expansion per random vector, the same as the diagonal elements alone. Second, we put forward a new local Chern marker: Using these maps we construct a local orbital magnetization built entirely from bond currents, whose chemical-potential derivative yields a local Chern marker that bypasses the position operator altogether. Because it is built from currents rather than from position-operator commutators, our marker obeys no zero-sum rule. Across an interface it interpolates between the bulk values instead of overshooting, and for a
homogeneous sample its average converges to the Chern number. It is independent of the choice of origin at any temperature and energy resolution, and its steps and extrema give the sense of circulation of the current.

For pedagogical reasons, the manuscript follows the order in which the two
results build on each other.
We first present the numerical improvement, and then the local Chern marker that it makes possible.
Section~\ref{sec:estimators} shows that the pair rotations used for the off-diagonal elements can be applied after the spectral filter, so that one set of random vectors gives all bond currents.
We then turn to the marker.
Section~\ref{sec:stream} formalizes the loop currents on a square lattice, where their construction is most transparent, and from them a local
orbital magnetization, whose chemical-potential derivative defines the local
Chern marker of Sec.~\ref{sec:marker}.
We test our findings on a system of noninteracting electrons on a disordered square lattice in the integer quantum Hall regime, both in the mobility gaps and at the center of a Landau level
(Sec.~\ref{sec:hall}).
We then extend the construction to a two-dimensional crystal,
i.e. a lattice with a general basis and hoppings of arbitrary range (Sec.~\ref{sec:general}), and apply it to a Chern domain in the Haldane model$\,$\cite{haldane1988}, where it is compared with the
projector marker at a domain wall (Sec.~\ref{sec:haldane}).

\section{Bond currents from one set of filtered vectors}
\label{sec:estimators}

The complete map of equilibrium local currents in tight-binding models constitutes the fundamental construction for the results presented in this work.
Each bond current is the imaginary part of an off-diagonal element of a function of the Hamiltonian, weighted by a hopping.
In Ref.~\cite{veiga2026markov}, each such element was obtained by rotating its disjoint pair of sites
into a basis where it appears as a difference of diagonal matrix elements.
This approach demanded a separate Chebyshev expansion for every tiling of the lattice into disjoint pairs, and additionally for the real and the imaginary parts.
We show here that one set of filtered random vectors gives every matrix element, and hence every bond current, while each current remains a difference of two non-negative samples whose fluctuations are set by the local spectral weight.

We consider a generic tight-binding Hamiltonian
\begin{equation}
H=\sum_{a,b}t_{ab}\,c^{\dagger}_{a}c_{b},\label{eq:H}
\end{equation}
where $c^\dagger_a$ is the creation operator for an atomic-like orbital $\varphi_a$ at position $\mathbf{r}_{a}$. Here, $t_{ab}$ corresponds to either a hopping amplitude between a pair of orbitals ($a\neq b$) or an on-site energy ($a=b$).
The information regarding typical equilibrium properties such as energy and magnetization is encoded in the matrix elements of positive semidefinite Hermitian spectral operators, $X=g(H)$, with $g\left(\dots\right)$
a real function of the Hamiltonian.
Important examples include the Fermi--Dirac density matrix $f(H)=\left[1+\exp\left(\beta\left(H-\mu\right)\right)\right]^{-1}$ at temperature $T=\left(k_{B}\beta\right)^{-1}$ and the spectral
window, $\delta_{\eta}(\varepsilon-H)$, a normalized Gaussian of standard deviation $\eta$ that broadens the Dirac delta.  Defining $X_{ab}=\langle a|X|b\rangle$ and $x_{a}=X_{aa}$, the equilibrium current of particles of charge
$e$ from $a$ to $b$ can be cast as
\begin{equation}
I_{ab}=\frac{2e}{\hbar}\,\mathrm{Tr}\big(XW_{ab}\big),\label{eq:current}
\end{equation}
where $W_{ab}=\tfrac{i}{2}\big(t_{ab}|a\rangle\langle b|-\mathrm{h.c.}\big)$ is the traceless Hermitian bond operator.

The stochastic estimation of the local current maps starts by generating a set of $R$ random vectors
\begin{equation}
|\xi_{n}\rangle=\sum_{a}e^{i\theta_{n,a}}|a\rangle,\qquad n=1,\dots,R,\label{eq:randvec}
\end{equation}
with independent phases drawn uniformly in $[0,2\pi)$. Since $\overline{\xi^{*}_{n,a}\xi_{n,b}}=\delta_{ab}$,
where the overline denotes the average over random vectors, we have that
\begin{equation}
\overline{|v_{n}\rangle\langle v_{n}|}=X,\label{eq:vv}
\end{equation}
with $|v_{n}\rangle=\sqrt{X}|\xi_{n}\rangle$.
From these filtered vectors, we can compute the full set of diagonal matrix elements, $x_{a}=\overline{|v_{n,a}|^{2}}$, as a mean of non-negative samples whose standard deviation is bounded by the mean$\,$\cite{veiga2026markov}.

Off-diagonal elements are estimated as differences of diagonal elements in a rotated basis. For a pair of orbitals $\{a,b\}$ and a phase $|\gamma|=1$, the states
\begin{equation}
|{+}_{\gamma}\rangle=\frac{|a\rangle+\gamma|b\rangle}{\sqrt{2}},\qquad|{-}_{\gamma}\rangle=\frac{-\gamma^{*}|a\rangle+|b\rangle}{\sqrt{2}}\label{eq:alphabeta}
\end{equation}
yield
\begin{equation}
\Upsilon_{\gamma}\equiv\langle{+}_{\gamma}|X|{+}_{\gamma}\rangle-\langle{-}_{\gamma}|X|{-}_{\gamma}\rangle=2\,\mathrm{Re}\big(\gamma X_{ab}\big).\label{eq:upsilon}
\end{equation}
In the original scheme, all elements of a tiling $\left(\cal T\right)$ of the lattice into disjoint pairs were rotated at once by a block-diagonal unitary $U_{\cal T}$, whose columns are the states $|{\pm}_{\gamma}\rangle$ of each pair $\left\{ a,b\right\} $.
The diagonal elements of $U^{\dagger}_{\cal T}XU_{\cal T}$ were sampled as in Eq.$\,$\eqref{eq:vv}, with $\sqrt{X}$ replaced by $U^{\dagger}_{\cal T}\sqrt{X}U_{\cal T}$.
The phases $\gamma=1$ and $\gamma=i$ give the real and imaginary parts of $X_{ab}$, and with $t_{ba}=|t_{ab}|e^{i\theta_{ba}}$ the current was assembled as
\begin{equation}
\frac{\hbar}{e}\,I_{ab}=|t_{ab}|\big(\sin\theta_{ba}\,\Upsilon_{1}-\cos\theta_{ba}\,\Upsilon_{i}\big).\label{eq:twopass}
\end{equation}
Each orbital belongs to exactly one pair of a tiling, so $z$ bonds per orbital require $N_{\cal T}\geq z$ tilings ($N_{\cal T}=4$ on the square lattice, $N_{\cal T}=9$ for the Haldane model).
As each tiling required its own expansion, for each random vector we had to perform $2N_{\cal T}$
independent simulations to obtain the complete map of local currents.
In the following paragraphs we show how this multiplicative cost can be eliminated exactly.

The first step consists of rotating into the eigenbasis of $W_{ab}$, rather than computing the real and imaginary parts of $X_{ab}$ independently.
On a particular pair, $\{a,b\}$, the structure of the local current is encoded in
\begin{equation}
W_{ab}=\frac{1}{2}\begin{pmatrix}0 & it_{ab}\\
-it_{ba} & 0
\end{pmatrix},\label{eq:W2}
\end{equation}
with eigenvalues $\pm|t_{ab}|/2$.
Its eigenvectors are the states of Eq.~\eqref{eq:alphabeta} evaluated at the bond-dependent phase
\begin{equation}
\gamma_{ab}=-i\,\frac{t_{ba}}{|t_{ab}|},\label{eq:gammaij}
\end{equation}
so that $W_{ab}=\tfrac{1}{2}|t_{ab}|\big(|{+}_{\gamma_{ab}}\rangle\langle{+}_{\gamma_{ab}}|-|{-}_{\gamma_{ab}}\rangle\langle{-}_{\gamma_{ab}}|\big)$.

\begin{figure}[tbp]
\includegraphics[width=0.9\columnwidth]{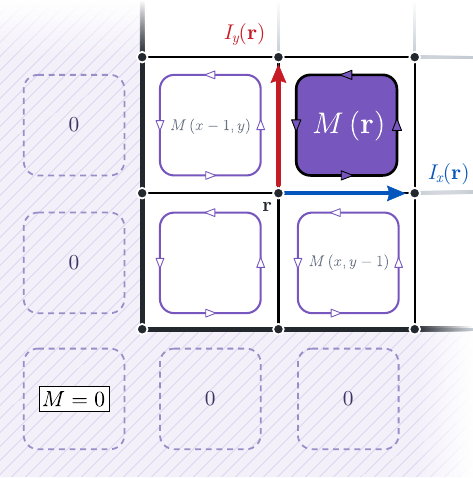}
\caption{Loop currents on the square lattice.
Each plaquette, labeled by its bottom-left corner $\mathbf{r}=(x,y)$, carries a loop current $M\left(\mathbf{r}\right)$ defined at the geometric center of the square.
Each bond current is the difference of the loop currents on its two sides: $I_{x}(\mathbf{r})=M(\mathbf{r})-M(x,y-1)$ and $I_{y}(\mathbf{r})=M(x-1,y)-M(\mathbf{r})$ {[}Eq.~\eqref{eq:square}{]}.
Outside the sample (shaded), $M=0$.}
\label{fig:stream}
\end{figure}

Using Eq.~\eqref{eq:upsilon} we obtain
\begin{equation}
I_{ab}=\frac{e}{\hbar}\,|t_{ab}|\,\Upsilon_{\gamma_{ab}}.\label{eq:singlepass}
\end{equation}
Since the pairs of a tiling are disjoint, this rotation with a bond-dependent phase can be applied simultaneously across the lattice, halving the number of expansions required by Eq.~\eqref{eq:twopass}.

The second step removes the multiplicative cost altogether.
It follows from comparing two estimators of the same rotated diagonal element $\langle u|X|u\rangle$, with $\ket u=U_{\cal T}\ket a$.
In the original scheme, the rotation acts twice, on both sides of $\sqrt{X}$, sampling the off-diagonal matrix elements contained in the tiling ${\cal T}$ through the estimator
\begin{equation}
\phi^{\left(\cal T\right)}_{n,u}=\bra{\xi_{n}}\circ U^{\dagger}_{\cal T}\sqrt{X}U_{\cal T}\ket{\xi_{n}},\label{eq:est2}
\end{equation}
with $\bra{\psi_{1}}\circ\ket{\psi_{2}}=\psi^{*}_{1}\left(a\right)\psi_{2}\left(a\right)$ denoting element-wise multiplication, whose component $a$ we label by the rotated state $\ket u$ that it samples.
Instead, the rotation can act once, after the spectral operator, on the same vector for every tiling,
\begin{equation}
\chi^{\left(\cal T\right)}_{n,u}=\bra{\xi_{n}}\circ U^{\dagger}_{\cal T}\sqrt{X}\ket{\xi_{n}}.
\label{eq:est1}
\end{equation}
For a fixed unitary transformation the estimators of Eqs.~\eqref{eq:est2} and \eqref{eq:est1} are both unbiased and converge to the same diagonal matrix element.
Since $U_{\cal T}$ is unitary and $\overline{U_{\cal T}\ket{\xi_{n}}\bra{\xi_{n}}U^{\dagger}_{\cal T}}=\openone$, we have that
\begin{equation}
\overline{\left|\phi^{(\cal T)}_{n,u}\right|^{2}}\,=\,\overline{\left|\chi^{(\cal T)}_{n,u}\right|^{2}},
\end{equation}
while the error bound associated with the diagonal estimator of Eq.~\eqref{eq:est2} carries over to $\chi^{(\cal T)}_{n,u}$.
This is easily verified by computing its variance
\begin{equation}
\mathrm{Var}\big(|\chi^{(\cal T)}_{n,u}|^{2}\big)=\langle u|X|u\rangle^{2}-\sum_{c}\big|\langle u|\sqrt{X}|c\rangle\big|^{4},\label{eq:variance}
\end{equation}
signifying that every diagonal matrix element in the rotated basis is again a mean of non-negative samples whose standard deviation does not exceed the mean.

The bond current, however, is the difference of two such non-negative samples drawn from the same random vector [Eq.~\eqref{eq:singlepass}].
While each rotated diagonal element retains the error bound of the diagonal estimator, the relative error of the current is not bounded in the same way, depending both on the model, and on the local observable being computed. This subtraction was already present in the original scheme. The gain is therefore entirely in computational efficiency.
For the computational time needed for
a single map in Ref.$\,$\cite{veiga2026markov}, which required $2N_{\cal T}$ expansions per random vector, the present scheme averages over $2N_{\cal T}$ times more random vectors, reducing the variance of every local current field by the same factor.

\section{Loop currents and a current-based Chern marker on the square lattice}
\label{sec:stream}

The systematic construction of the loop current field begins with the Helmholtz decomposition of a planar current field,
\begin{equation}
\mathbf{I}=\mathbf{I}_{{\rm div}}+\mathbf{I}_{{\rm circ}}+\mathbf{I}_{{\rm tr}}.\label{eq:helmholtz}
\end{equation}
Here $\mathbf{I}_{{\rm div}}=\nabla\zeta$, with $\nabla^{2}\zeta=\nabla\cdot\mathbf{I}$, carries the sources and sinks, $\mathbf{I}_{{\rm circ}}=\nabla\times(M\,\mathbf{e}_{z})$ the circulation, and the harmonic component $\mathbf{I}_{{\rm tr}}=\nabla\upsilon$, with $\nabla^{2}\upsilon=0$, a transport current. Equilibrium currents are steady, $\mathbf{I}_{{\rm div}}=0$, and carry no net transport, $\mathbf{I}_{{\rm tr}}=0$, so that
\begin{equation}
\mathbf{I}=\nabla\times(M\,\mathbf{e}_{z}),
\label{eq:curl_current}
\end{equation}
or equivalently
\begin{equation}
(\nabla\times\mathbf{I})_{z}=-\nabla^{2}M,\label{eq:poisson}
\end{equation}
with $M=0$ outside a sample with open boundaries. On a tight-binding lattice, charge is defined on the nodes, while local currents flow through the connecting bonds.
On the square lattice we define loop currents $M(\mathbf{r})$ on the plaquettes, each labeled by its bottom-left corner $\mathbf{r}=(x,y)$.

Each bond current is the difference of the loop currents of the two plaquettes that share the
bond, the lattice form of Amp\`ere's law~\cite{nathan2017quantized} (Fig.~\ref{fig:stream}),
\begin{equation}
\begin{aligned}
I_{x}(\mathbf{r}) & =M(\mathbf{r})-M(x,y-1),\\
I_{y}(\mathbf{r}) & =M(x-1,y)-M(\mathbf{r}),
\end{aligned}\label{eq:square}
\end{equation}
with $I_{x}(\mathbf{r})$ and $I_{y}(\mathbf{r})$ being the currents from $\mathbf{r}$ to $\mathbf{r}+\mathbf{e}_{x}$ and to $\mathbf{r}+\mathbf{e}_{y}$, respectively. Equation~\eqref{eq:square} can be solved exactly only if the current is conserved at every site, which is also the condition for the loop currents to be independent of the path along which Eq.~\eqref{eq:square} is integrated~\cite{nathan2017quantized},
\begin{equation}
\sum_{b}I_{ab}[X]=0,\label{eq:conservation}
\end{equation}
which represents the lattice statement of $\mathbf{I}_{\rm div}=0$.
In that case the loop current is obtained directly as
\begin{equation}
M(\mathbf{r})=\sum_{y'\leq y}I_{x}(x,y'),\label{eq:recursion}
\end{equation}
or equivalently, by Eq.~\eqref{eq:conservation},
\begin{equation}
M(\mathbf{r})=-\sum_{x'\leq x}I_{y}(x',y).\label{eq:recursion1}
\end{equation}
Summing by parts with $M=0$ outside the sample gives
\begin{equation}
\int M\,d^{2}r=\frac{1}{2}\int(\mathbf{r}\times\mathbf{I})_{z}\,d^{2}r,\label{eq:summation_m}
\end{equation}
which corresponds to the orbital magnetization of a finite sample \cite{walz2014, bahamon2020, vidarte2026}.
Hence $M$ is a local orbital magnetization.
Unlike the integrand $\tfrac{1}{2}(\mathbf{r}\times\mathbf{I})_{z}$, which changes under a shift of origin and becomes origin-independent only once integrated, $M(\mathbf{r})$ is fixed by the bond currents alone.
It involves no position operator, and its local definition is independent of the choice of origin. However, stochastic currents are conserved only on average
rendering Eq.~\eqref{eq:square}
overdetermined.
We solve it using least-squares minimization, whose normal equations
are exactly the lattice form of Eq.~\eqref{eq:poisson}, with the vorticity of the local current field taking the role of the source term.

\begin{figure}[t]
\includegraphics[width=1\columnwidth]{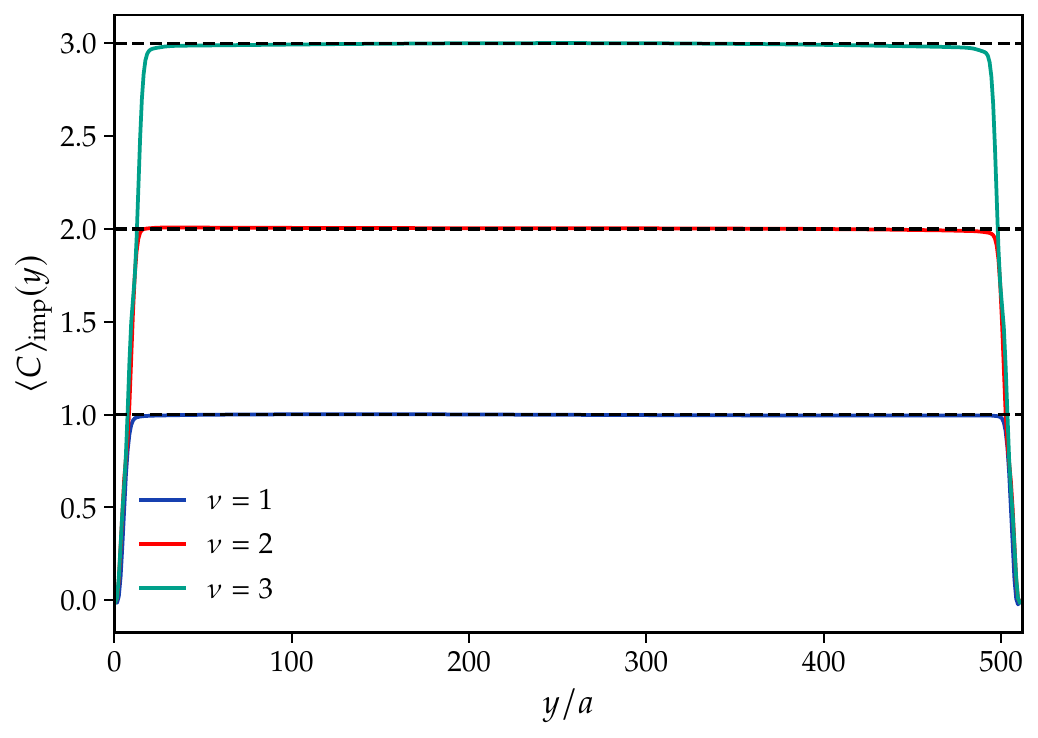}
\caption{Profile of the current marker across an open $512\times512$ square
lattice in the presence of a perpendicular-to-plane magnetic field, with flux $\Phi=1/100$ per plaquette,
and Anderson disorder (on-site energies drawn uniformly from $[-W/2,W/2]$, with
$W=0.25\,t$).
The chemical potentials $\mu=-3.88\,t$, $-3.75\,t$ and $-3.635\,t$ lie in the
gaps above the first three Landau levels, $\nu=1,2,3$ ($\mu_{a}$ and $\mu_{c}$
of Fig.~\ref{fig:hall}).
Solid lines: $\imp{C}$, the marker $C_{\eta}$ ($\eta=5\times10^{-3}\,t$)
averaged over 16 disorder configurations, along the column $x=L/2$, as a
function of $y$.
Horizontal dashed lines: $C=1,2,3$. In the bulk the marker is flat at the
filling $\nu$, to within $1\%$, while at the edges it falls to zero over a few
magnetic lengths, $\ell_{B}=a/\sqrt{2\pi\Phi}\approx4a$.}
\label{fig:hallcut}
\end{figure}

\begin{figure*}[t]
\includegraphics[width=\textwidth]{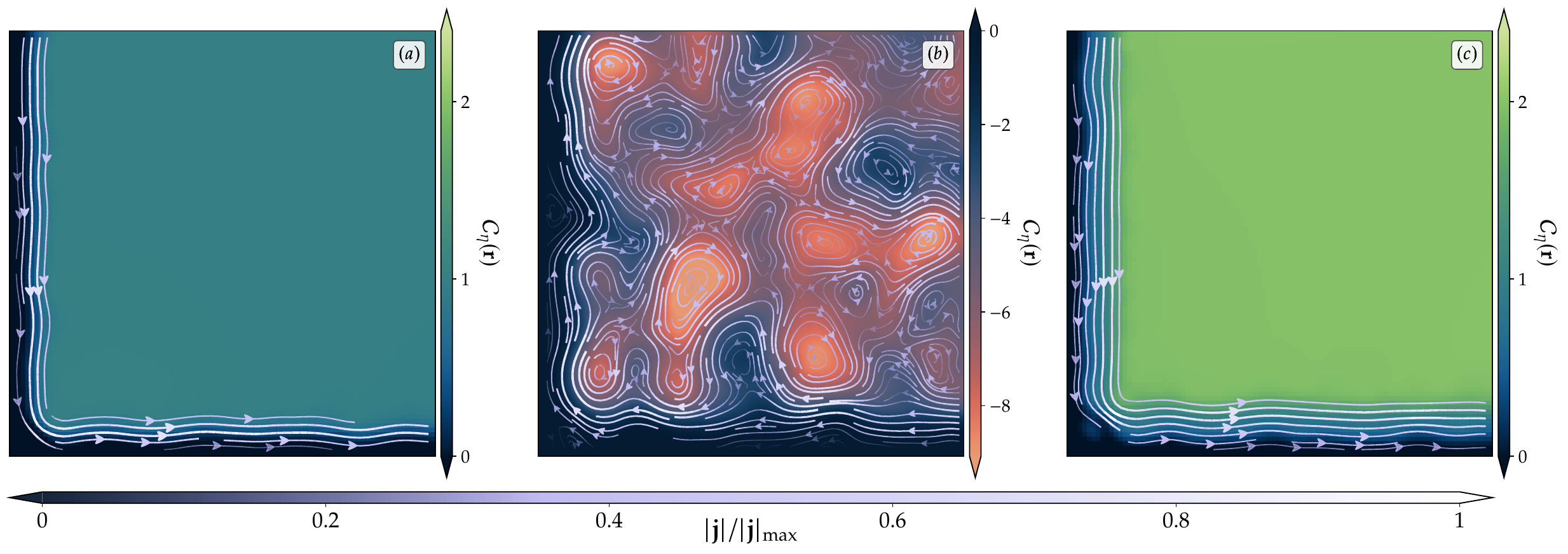}
\caption{Current marker and spectral currents in an open $512\times512$ square lattice
in a perpendicular magnetic field, with flux $\Phi=1/100$ per plaquette and
Anderson disorder $W=0.25\,t$, for a single disorder realization and three
chemical potentials:
(a) $\mu=-3.88\,t$, in the gap above the lowest Landau level, $E_{0}<\mu<E_{1}$. In
(b) $\mu=-3.82\,t\simeq E_{1}$, at the center of the second Landau level, while for 
(c) $\mu=-3.75\,t$, in the gap above it, $E_{1}<\mu<E_{2}$.
Each panel shows the same $96\times96$ window at the bottom-left corner of the lattice.
Background: $C_{\eta}(\mathbf{r})$, lighter for larger $|C_{\eta}|$. In (a) and (c)
it is shown site by site on a shared color scale from $0$ to $2.4$. In (b) it is
averaged over the same $4\times4$ blocks as the currents, on a scale running from
the 1st percentile of the window to $0$. Values outside a scale take its end colors.
Lines: streamlines of the spectral current, averaged over $4\times4$
blocks, with color and width set by $|\mathbf{j}|$ relative to its
maximum in each panel (currents below $1\%$ of the maximum are not
shown). On the plateaus the current flows only along the edges,
counter-clockwise around the region of positive marker, and the bulk
is free of current.
At the center of the level the current penetrates the whole window:
the bright spots of (b), local minima of $C_{\eta}$, are encircled by
clockwise whirlpools, and the darker regions between them, local maxima,
by counter-clockwise ones.}
\label{fig:hall}
\end{figure*}

\subsection{A current-based Chern marker}

\label{sec:marker}
\label{sec:markers}

When the chemical potential $\mu$ lies within the bulk gap of an insulator, the Středa relation~\cite{streda1982,macdonald1984} connects the global orbital magnetization to the Chern number $C$,
\begin{equation}
\frac{\partial M_{z}}{\partial\mu}=\frac{eA}{h}\,C,\label{eq:streda}
\end{equation}
where $A$ is the area of the sample and $C\equiv(h/e^{2})\,\sigma_{xy}$, with $j_{x}=\sigma_{xy}E_{y}$, being the Hall conductivity in units of $e^{2}/h$ \footnote{This $C$ is minus the Berry-curvature Chern number $C_{\rm B}=(2\pi)^{-1}\int\Omega\,d^{2}k$ of~\cite{thonhauser2005,ceresoli2006}. For electrons, $e=-|e|$, Eq.~\eqref{eq:streda} takes the familiar form $\partial M_{z}/\partial\mu=|e|A/h\,\,C_{\rm B}$.}. Since $M_{z}=\int M\,d^{2}r$ [Eq.~\eqref{eq:summation_m}], a local version of this relation follows
from the chemical-potential derivative of the loop current.
Writing the single-particle density matrix as
\begin{equation}
\rho(H)=\int d\varepsilon\,f(\varepsilon-\mu)\,\delta(\varepsilon-H),
\end{equation}
where $f$ denotes the Fermi--Dirac occupation, every bond current decomposes into individual energy contributions, $I_{ab}=\int d\varepsilon\,f(\varepsilon-\mu)\,j_{ab}(\varepsilon)$, where the spectral bond current
is given by
\begin{equation}
j_{ab}(\varepsilon)=\frac{2e}{\hbar}\,\mathrm{Im}\big[t_{ba}\,\delta(\varepsilon-H)_{ab}\big].\label{eq:spectral}
\end{equation}
By linearity, the spectral currents obey the same continuity equation as $I_{ab}$, and Eq.~\eqref{eq:recursion} applied to them defines the spectral loop current $m(\mathbf{r},\varepsilon)$ satisfying
\begin{equation}
M(\mathbf{r})=\int d\varepsilon\,f(\varepsilon-\mu)\,m(\mathbf{r},\varepsilon)
\end{equation}
and, by Eq.~\eqref{eq:recursion}, we have that
\begin{equation}
m(\mathbf{r},\varepsilon)=\sum_{y'\leq y}j_{x}(x,y';\varepsilon).
\end{equation}
At zero temperature $\partial\rho/\partial\mu=\delta(\mu-H)$, hence
\begin{equation}
\frac{\partial M(\mathbf{r})}{\partial\mu}=m(\mathbf{r},\mu),\label{eq:dMdmu}
\end{equation}
which implies that the derivative of the local magnetization is the
spectral loop current
at the Fermi level.
Integrated over the sample it yields $\partial M_{z}/\partial\mu$.
Consequently, Eq.~\eqref{eq:dMdmu} is the local form of Eq.~\eqref{eq:streda}. We therefore define a local Chern marker
\begin{equation}
C_{\eta}(\mathbf{r},\mu)\equiv\frac{h}{e}\int d\varepsilon\,\delta_{\eta}(\mu-\varepsilon)\,m(\mathbf{r},\varepsilon),\label{eq:localC}
\end{equation}
from the spectral loop current resolved within an energy resolution  $\eta$ around the Fermi
level, in units of $e/h$.  Setting $\delta_{\eta}=-f'$  yields $(h/e)\,\partial M/\partial\mu$ at the inverse temperature $\beta$. In a finite sample the spectrum is discrete, and inside a gap the only states near $\mu$
belong to the chiral edge channels, with level spacing $\delta\varepsilon\simeq2\pi\hbar v/\ell$
for a boundary of length $\ell$ and edge velocity $v$. The marker is therefore defined in the regime
\begin{equation}
\delta\varepsilon\lesssim\eta\ll\Delta,\label{eq:window}
\end{equation}
where the window contains many edge levels, so that the edge spectrum acts as a continuum,
while excluding the bulk bands across the gap $\Delta$. This is the same order of limits,
$\ell\to\infty$ before $\eta\to0$, under which the Středa relation~\eqref{eq:streda} holds.
In the opposite limit, $\eta\ll\delta\varepsilon$ with $\mu$ between two levels,
$\delta_{\eta}(\mu-H)$ vanishes and so does $C_{\eta}$, everywhere in the sample. Using Eq.~\eqref{eq:curl_current}, we find that
\begin{equation}
\mathbf{j}_{\eta}=(e/h)\,\nabla\times(C_{\eta}\,\mathbf{e}_{z}),
\end{equation}
where $\mathbf{j}_{\eta}=\int d\varepsilon\,\delta_{\eta}(\mu-\varepsilon)\,\mathbf{j}(\varepsilon)$ is the spectral current resolved within $\eta$, the field plotted in Fig.~\ref{fig:hall}.
Therefore, a step $\Delta C$ carries the net spectral current $\Delta C\,e/h$ along it, and the spectral current circulates counter-clockwise around maxima of $C_{\eta}$ and clockwise around minima, so the sense of rotation is given by the sign of the step for an edge channel, and by the type of extremum for a vortex.

Built from bond currents alone, ${C}_{\eta}$ involves no position operator and is independent of the origin for any $\eta$ and temperature. Projector markers share this property only when $P$ is an exact projector. At finite temperature, $P^{2}\neq P$ at metallic edges and walls, and a shift of origin changes the values acquired by the marker.

\section{Current vortices in a quantum Hall lattice}
\label{sec:hall}

The integer quantum Hall effect has been the paramount example of a topological system.
In fact, it is where equilibrium currents, magnetization, and topology were first linked~\cite{streda1982,macdonald1984,geller1994} and, recently, where equilibrium currents have been imaged~\cite{uri2020}.
While the construction detailed in Sec.~\ref{sec:stream} applies directly to the tight-binding square lattice, this system tests the marker in two regimes.
On the one hand, in a mobility gap, it should reproduce the Chern number corresponding to each topological gap.
On the other hand, inside a disorder-broadened Landau level, the spectral current invades the system's bulk and the marker is no longer pinned to an integer value.
We take the Hofstadter model~\cite{hofstadter1976}, a square lattice with hopping $-t$ and
magnetic flux $\Phi=1/100$ per plaquette (in units of the magnetic flux quantum $\Phi_{0}$) in the Landau gauge, on a $512\times512$ flake ($2.6\times10^{5}$ sites) with Anderson disorder.
Each uncorrelated on-site energy is drawn from a uniform probability distribution defined on $[-W/2,W/2]$, with $W=0.25\,t$.
The three lowest Landau levels of the clean lattice peak at $E_{0}\simeq-3.94 t$, $E_{1}\simeq-3.81 t$, and $E_{2}\simeq-3.69 t$, while the cyclotron gap is $\hbar\omega_{c}\simeq0.12\,t$.
We consider three distinct chemical potentials: $\mu_{a}=-3.88\,t$ and $\mu_{c}=-3.75\,t$, in the mobility gaps below and above the second level, and $\mu_{b}=-3.82\,t$ which lies within the center of the second Landau level.
For the profiles, we also consider $\mu=-3.635 t$, situated in the gap above the third level.
We take $\eta=5\times10^{-3}\,t$, larger than the spacing of the edge levels, and $R=2048$ random vectors per $\mu$ and disorder configuration. The maps of Fig.~\ref{fig:hall} are for a
single disorder realization, while the profiles of Fig.~\ref{fig:hallcut} are averaged over 16 independent configurations.

\begin{figure}[tbp]
\includegraphics[width=0.9\columnwidth]{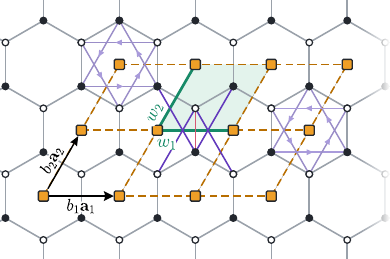}\caption{Coarse cells of size $b_{i}=1$ on the Haldane lattice. Parallelograms
are spanned by $b_{1}\mathbf{a}_{1}$ and $b_{2}\mathbf{a}_{2}$ (dashed),
with corners (squares) on hexagon centers. The colored bonds cross
the bottom wall $w_{1}$: one nearest-neighbor bond and four next-nearest-neighbor
chords.}
\label{fig:coarse}
\end{figure}

In the mobility gap above the lowest level the local marker is uniform in the bulk, rendering the Chern number $C = 1$ of the plateau, and above the second and third levels it is $C = 2$ and $C = 3$, respectively
{[}Fig.~\ref{fig:hallcut}{]}. This is the bulk value expected for the projector marker~\cite{kitaev2006,bianco2011}
\begin{equation}
C_{\rm P}(\mathbf{r}_{a})=\frac{4\pi}{A_{s}}\,\mathrm{Im}\langle a|PxQyP|a\rangle,\label{eq:BR}
\end{equation}
with $P$ the ground-state projector, $Q=\openone-P$ its complement and $A_{s}$ the area per site.
That $C_{\eta}$ reproduces this value follows from the edge currents.
An in-gap window carries spectral current only along the boundary, where each chiral channel carries the jump of the Chern number across the gap \cite{hatsugai1993,kellendonk2002,elgart2005}.
Along the cut these jumps add up from $0$ outside the sample to $C$, with corrections exponentially small in the distance to the nearest boundary and of order $e^{-\Delta^{2}/8\eta^{2}}$ for a gap $\Delta$~\cite{elgart2005,prodan2016}.
The window must, however, resolve the discrete edge spectrum, which has a level spacing of $\delta\varepsilon\simeq2\pi\hbar v/\ell$ for a boundary of length $\ell$.
By Poisson summation the relative error is $2e^{-2\pi^{2}\eta^{2}/\delta\varepsilon^{2}}$, meaning the marker requires $\delta\varepsilon\lesssim\eta\ll\Delta$, a condition our choice of $\eta$
satisfies.

Towards the edges, however,
the local markers $C_\eta$ and $C_{\rm P}$ exhibit distinct behaviors.
$C_{\eta}$ falls to zero over a few magnetic lengths {[}Fig.~\ref{fig:hallcut}{]},
corresponding to the spatial region
where the spectral current flows {[}Fig.~\ref{fig:hall}(a) and \ref{fig:hall}(c){]}.
It runs down the left edge and to the right along the bottom edge, counter-clockwise around the sample, as dictated by the positive sign of the marker
(Sec.~\ref{sec:marker}), and carries the net spectral current $C\,e/h$, which is twice as large in (c) as in (a).
The projector marker $C_{\rm P}$ must instead overshoot at the edge since, for an exact projector, its sum over the sample vanishes,
\begin{align}
\sum_{a}A_{s}\,C_{{\rm P}}(\mathbf{r}_{a}) & =0,\label{eq:sumrules}
\end{align}
which forces its bulk value to be compensated at the boundaries.
Conversely, for
the current-based local Chern marker  $C_\eta$, the sample average converges to the Chern number with finite-size corrections of order the perimeter times the channel depth divided by the area.

The projector marker is a local quantity when the ground-state projector $P$ decays exponentially in real space.
This occurs whenever $\mu$ lies in a gap or a mobility gap~\cite{kohn1996,prodan2016}.
At the center of a Landau level, the localization length diverges~\cite{huckestein1995,evers2008} and $P$ decays as a power law.
$C_{\rm P}(\mathbf{r}_{a})$ then collects contributions from distant regions of the sample, and its fluctuations and correlation length grow as the transition is approached~\cite{favata2025}.
The current marker has no such requirement.
Its input is the local current, a nearest-neighbor matrix element of the smooth window
$\delta_{\eta}(\mu-H)$ defined at every energy, so $C_{\eta}$ remains a local quantity throughout the band.
Its value, however, is no longer pinned to an integer, and its sum over the whole sample equals $(h/e)\,\partial M_{z}/\partial\mu$, carried by the states within $\eta$ of $\mu$. Within the window of
Fig.~\ref{fig:hall}(b) it varies on the scale $\ell_{B}$, dominated by compact local minima (bright spots).
Since $C_{\eta}$ is the stream function of the spectral current,
its local extrema determine the direction of circulation. The strong whirlpools around the minima rotate clockwise, while weaker whirlpools around local maxima rotate counter-clockwise, and between them the current follows open, meandering channels.

\section{Generalization to arbitrary two-dimensional crystals}
\label{sec:general}

On the square lattice with nearest-neighbor hopping, every bond separates two plaquettes, and Eq.~\eqref{eq:square} assigns a loop current to each of them.
For a general model, such as a lattice with a multi-site, multi-orbital basis and
arbitrary-range hoppings, this approach is no longer convenient.
Bonds drawn as straight segments intersect one another and divide the plane into plaquettes of
diverse shapes.
We therefore do not resolve the loop currents on these plaquettes.
Instead, we evaluate them on a regular grid of points defined by the primitive vectors $\mathbf{a}_1$, $\mathbf{a}_2$ of the underlying Bravais lattice, thereby extending the construction of Sec.~\ref{sec:stream} to any two-dimensional crystal.

For integers $b_{1},b_{2}\geq1$, the coarse cells are the parallelograms
spanned by $b_{1}\mathbf{a}_{1}$ and $b_{2}\mathbf{a}_{2}$, with corners
\begin{equation}
\mathbf{c}_{N_{1}N_{2}}=\mathbf{c}_{0}+b_{1}N_{1}\mathbf{a}_{1}+b_{2}N_{2}\mathbf{a}_{2},\label{eq:cornerpos}
\end{equation}
so that $b_{1}=b_{2}=1$ recovers the unit cells.
The offset $\mathbf{c}_{0}$ is chosen so that no site lies on a cell edge.
For instance, on the honeycomb lattice of the Haldane model, the corners
can be chosen as the hexagon centers (see Fig.~\ref{fig:coarse}).
Each cell owns the two walls $w_{1}$ and $w_{2}$ that extend from its corner $\mathbf{c}_{N_{1}N_{2}}$ along $b_{1}$$\mathbf{a}_{1}$ and $b_{2}$$\mathbf{a}_{2}$.
The net current through a wall,
\begin{equation}
\cal I_{k}\left(N_{1},N_{2}\right)=\sum_{\langle a,b\rangle}\sigma_{ab}\big(w_{k}\big)\,I_{ab},\label{eq:wallflux}
\end{equation}
collects only the bonds that cross it, where $\sigma_{ab}(w_{k})=\pm1$ is positive when
the bond $a\to b$ crosses $w_{k}$ from left to right, as seen walking along $w_{k}$ away from its corner, and negative otherwise. For instance, in the Haldane model  with $b_{i}=1$ (discussed below), each wall is crossed by one nearest-neighbor bond and four next-nearest-neighbor chords
(Fig.~\ref{fig:coarse}).

Analogous to the square lattice construction, the loop current is constant between bonds and jumps by $I_{ab}$ across the bond $a\to b$.
Following a wall from one corner to the next therefore accumulates the jumps of all the bonds that cross it, and the corner loop currents $M_{N_{1}N_{2}}\equiv M(\mathbf{c}_{N_{1}N_{2}})$
obey
\begin{align}
M_{N_{1}+1,N_{2}}-M_{N_{1}N_{2}} & =\cal I_{1}\left(N_{1},N_{2}\right),\nonumber \\
M_{N_{1},N_{2}+1}-M_{N_{1}N_{2}} & =\cal I_{2}\left(N_{1},N_{2}\right),\label{eq:corners}
\end{align}
with $M=0$ on the corners outside the sample. Going counter-clockwise around a cell, the wall currents add up to the net current leaving the sites it encloses, which vanishes by Eq.~\eqref{eq:conservation}.
Equation~\eqref{eq:corners} is therefore consistent, and $M_{N_{1}N_{2}}$ does not depend on the path along which the walls are summed.
On the square lattice with $b_{1}=b_{2}=1$ and $\mathbf{c}_{0}$ at a plaquette center, each wall is crossed by a single bond, and Eq.~\eqref{eq:corners} reduces to Eq.~\eqref{eq:square}.

As on the square lattice, stochastic currents are conserved only on average,
and we determine the loop current field
using least-squares minimization.
Accordingly, when
applied to the spectral currents $j_{ab}(\varepsilon)$, the same construction gives the spectral loop currents $m_{N_{1}N_{2}}(\varepsilon)$.
Through Eq.~\eqref{eq:localC}, these quantities define the local Chern marker $C_{\eta}(\mathbf{c}_{N_{1}N_{2}},\mu)$ at the corners of the coarse cells.

\begin{figure}[tbp]
\includegraphics[width=1\columnwidth]{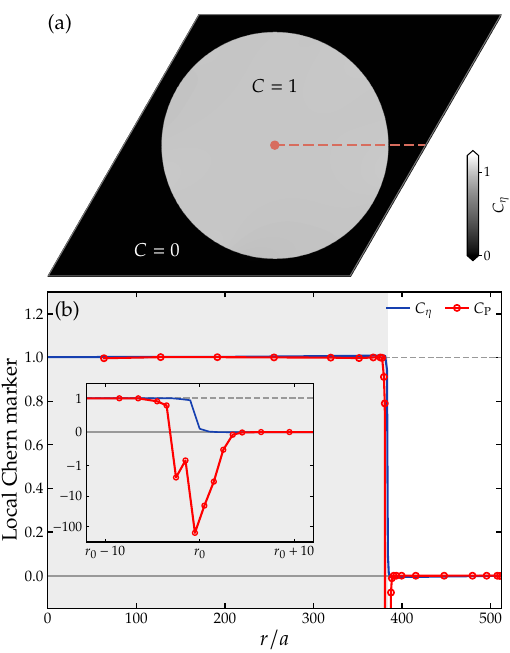}
\caption{Local Chern markers for a Chern domain ($\epsilon_{{\rm s}}=0$ for $r<r_{0}=384a$)
embedded in a trivial Haldane insulator ($\epsilon_{{\rm s}}=3t_{1}$) on a $1024\times1024$ flake with disorder $W=0.5\,t_{1}$, at $\mu=0$.
(a) Current marker $C_{\eta}$ on the whole flake, from one set of $R=4096$ random vectors;
the dashed line indicates the cut shown in panel (b).
(b) Cut along the row $n_{2}=L/2$ from the center of the domain to the edge; shaded region: the domain $r<r_{0}$.
Solid line: current marker at hexagon centers.
Circles: projector marker, evaluated deterministically cell by cell.
Inset: details of the wall region, $r_{0}\pm12a$, on a symmetric-log scale. The projector marker overshoots to about $-160$ just inside the wall, while the current marker passes between the bulk values over a few cells.}
\label{fig:haldane}
\end{figure}

\section{Chern domain in the Haldane model}
\label{sec:haldane}

We demonstrate the effectiveness of
the construction of Sec.~\ref{sec:general} on the Haldane model \cite{haldane1988}, featuring a topological domain at the center of a trivial insulator.
Unlike the quantum Hall lattice of Sec.~\ref{sec:hall}, this is a Chern insulator without a net magnetic field, whose crossing next-nearest-neighbor hoppings require the coarse cells.
A Chern domain is also the simplest setting in which the current and projector markers must differ inside the sample, at the domain wall, and not only at its outer edge.
Such domains occur in magnetic topological insulators \cite{yasuda2017}, twisted graphene multilayers \cite{polshyn2020,tschirhart2021,grover2022}, and ultracold-atom realizations of the Haldane model \cite{jotzu2014}.

The Hamiltonian is
\begin{equation}
H=-t_{1}\sum_{\langle a,b\rangle}c_{a}^{\dagger}c_{b}
+t_{2}\sum_{\langle\langle a,b\rangle\rangle}e^{i\nu_{ab}\phi}\,c_{a}^{\dagger}c_{b}
+\sum_{a}\epsilon_{a}\,c_{a}^{\dagger}c_{a},
\end{equation}
where $\nu_{ab}=\mathrm{sgn}\big[(\mathbf{r}_{k}-\mathbf{r}_{b})\times(\mathbf{r}_{a}-\mathbf{r}_{k})\big]_{z}$,
with $k$ the common nearest neighbor of $a$ and $b$, i.e.\ $\nu_{ab}=+1$ when an electron
hopping from $b$ to $a$ turns counter-clockwise, $t_{2}=t_{1}/3$ and $\phi=\pi/2$.
The on-site energy $\epsilon_{a}=\epsilon_{s}(\mathbf{r}_{a})\,s_{a}+V_{a}$ combines a
staggered potential, with $s_{a}=+1$ ($-1$) on sublattice A (B), and Anderson disorder
$V_{a}$ drawn uniformly in $[-W/2,W/2]$ with $W=0.5\,t_{1}$. The pristine model is a Chern insulator for $|\epsilon_{{\rm s}}|<\sqrt{3}\,t_{1}$.
The sample is an $L\times L$ flake with open boundaries, consisting of $L=1024$ unit cells per side ($2.1\times10^{6}$ orbitals).
A disk of radius $r_{0}=384a$, with $a=|\mathbf{a}_{1}|$ the lattice constant, is defined by
setting $\epsilon_{\rm s}=0$ for sites at distance $r<r_{0}$ from the center of the flake
(yielding a Chern insulator with band edges at $\pm t_{1}$), with a sharp step to the
topologically trivial background, $\epsilon_{\rm s}=3t_{1}$.

The two markers are expected to behave differently at the wall.
The projector marker sums to zero over the sample [Eq.~\eqref{eq:sumrules}], requiring it to compensate for the bulk contribution of the domain within an annulus along the wall
whose width $w$ is set by the decay length of $P$.
In this region, it reaches values of order $-Cr_{0}/2w$, which grow with the size of the domain; for $w\approx1.2a$ this gives $\approx-160$, the value found in Fig.~\ref{fig:haldane}. The current marker is subject to no such constraint.
It interpolates between the bulk values, and its step carries the current $C\,e/h$ along the wall, counter-clockwise around the domain for $C>0$.

Both markers are computed using Chebyshev expansions in KITE \cite{joao2020kite,Pires2022,Castro_2024}, for the same domain geometry. The current marker is mapped across the whole flake from a single set of 4096 random vectors and
$\eta=1\times10^{-2} t_1$. The projector marker, although deterministic, requires two expansions per orbital, one for $P|a\rangle$ and another for $QyP|a\rangle$, where $P$ is a Fermi function at $\beta=2048\,t^{-1}_{1}$ and positions are measured from the center of the domain.
Since a map of the whole flake would require $4L^{2}$ such expansions, we evaluate
the projector marker on 23 cells of the row $n_{2}=L/2$, from the center of the disk to the edge, with spacing decreasing from $r_{0}/4$ in the bulk to one cell across the wall. Figure~\ref{fig:haldane} shows this comparison.
Deep inside both bulk regions the two markers agree with the bulk Chern numbers, $C=1$ and $C=0$.
At the wall, however, the projector marker overshoots.
The current marker instead interpolates smoothly between the bulk values over a few cells, corresponding to the width of the wall mode. Its statistical error decreases as $R^{-1/2}$, and $\eta$ lies in the regime $\delta\varepsilon\lesssim\eta\ll\Delta$ of Eq.~\eqref{eq:window}.

\section{Conclusions}

Our work establishes two main results.
First, applying pair rotations after the spectral filter yields all bond currents from a single set of random vectors.
This
eliminates the $2N_{\cal T}$ expansions per random vector required in Ref.~\cite{veiga2026markov}, thereby reducing the variance of each current by the same factor at equivalent computational cost.
Second, current conservation maps these currents into loop currents,
defined
at the corners of coarse cells in arbitrary two-dimensional crystals, which, for the Fermi sea, correspond to a local orbital magnetization.
Their derivative with respect to the chemical potential defines a local Chern marker that requires no position operator and no compensation at boundaries.
In a disordered quantum Hall lattice it
recovers the Chern number of each plateau, while at the center of a Landau level its contours trace current vortices whose sense of rotation is set by whether they encircle maxima or minima of the marker.
At a Chern domain wall in the Haldane model, it transitions between the bulk values over the width of the wall mode, where the projector marker overshoots to values that are expected to grow linearly with the
radius of the domain.

In incompressible regions, the marker, proportional to $\partial M/\partial\mu$, is the quantity imaged by gate-modulated scanning magnetometry \cite{tschirhart2021,grover2022}, whereas in compressible regions the measured signal corresponds to $\partial M/\partial n$.
The coarse cells can be enlarged ($b_{i}\gg1$) to the resolution set by the tip height, and convolving $M$ with the point-spread function of the magnetometer turns the maps of Secs.~\ref{sec:hall} and~\ref{sec:haldane} into predictions for such measurements.
Two natural extensions are the subject of ongoing work.
At the quantum Hall transition the loop currents form a random surface whose contours percolate, and it remains to be seen whether the spectral currents share the multifractality of the wave functions \cite{evers2008,mu2026}.
Furthermore, this estimator also applies to Bogoliubov--de Gennes Hamiltonians, where the
same stochastic vectors give the pairing amplitudes and supercurrents.

\begin{acknowledgments}
H.P.V. thanks J. M. Alendouro Pinho for fruitful discussions.
H.P.V. and J.M.V.P.L. acknowledge support from Funda\c{c}\~{a}o para a Ci\^{e}ncia e a Tecnologia (FCT, Portugal) through the Strategic Funding UID/04650/2025-Centro de F\'{i}sica das Universidades do Minho e do Porto. H.P.V. further acknowledges support from FCT through Project No.~EXPL/FIS-MAC/0953/2021 and Grant No.~2024.00560.BD. C.H.L. acknowledges support from the Deutsche Forschungsgemeinschaft (DFG, German Research Foundation) within Project-ID 314695032 -- SFB 1277, and from the Brazilian funding agencies FAPERJ and CNPq. H.P.V. acknowledges computational resources from the Laboratory for Advanced Computing at the University of Coimbra. A.F., J.M.V.P.L. and H.P.V. acknowledge computational resources from the University of York High-Performance Computing service, Viking.
\end{acknowledgments}

\bibliography{refs}

\end{document}